\documentclass[10pt,conference]{IEEEtran}
\usepackage{epsfig,setspace,amsmath,epsf,amssymb,bm,theorem,cite,graphicx, epstopdf,algorithm,float,color,mathtools, authblk}
\usepackage[table,xcdraw]{xcolor}
\usepackage{subcaption}
\usepackage{algpseudocode}
\usepackage{dsfont}

\DeclareMathOperator*{\argmax}{arg\,max}

\newtheorem{theorem}{Theorem}
\newtheorem{corollary}{Corollary}
\newtheorem{definition}{Definition}
\newtheorem{remark}{Remark}
\newtheorem{lemma}{Lemma}

\newcommand{\e}{{\mathbb{E}}}

\IEEEoverridecommandlockouts
\allowdisplaybreaks

\begin{document}

\title{Structured Estimators: A New Perspective on Information Freshness}

\author[1]{Sahan Liyanaarachchi}
\author[1]{Sennur Ulukus}
\author[2]{Nail Akar}

\affil[1]{\normalsize University of Maryland, College Park, MD, USA}
\affil[2]{\normalsize Bilkent University, Ankara, T\"{u}rkiye}

\maketitle
\begin{abstract}
    In recent literature, when modeling for information freshness in remote estimation settings, estimators have been mainly restricted to the class of martingale estimators, meaning the remote estimate at any time is equal to the most recently received update. This is mainly due to its simplicity and ease of analysis. However, these martingale estimators are far from optimal in some cases, especially in pull-based update systems. For such systems, maximum aposteriori probability (MAP) estimators are optimum, but can be challenging to analyze. Here, we introduce a new class of estimators, called structured estimators, which retain useful characteristics from a MAP estimate while still being analytically tractable. Our proposed estimators move seamlessly from a martingale estimator to a MAP estimator. 
\end{abstract}

\section{Introduction}
In many remote estimation applications, timely estimation of the underlying system state can be critical, especially in scenarios where crucial control decisions must be taken based on the current estimates of the system. In such applications, age of information (AoI) is considered the prominent metric to quantify the timeliness of the estimates \cite{yates2020age, age1, age2}. However, recent studies have shifted the focus to its more versatile counterpart known as the age of incorrect information (AoII) \cite{AoII2019, ismail1, ismail2}. AoI metric penalizes the staleness of the updates linearly as the time from the last update increases, whereas AoII metric penalizes the system only if there is a disparity between the actual system state and its current estimate. Hence, AoII is more befitting to the remote estimation setting.

Majority of the remote estimation applications model their underlying system as a Markovian source \cite{subhankar, Markov_machines}. Among the many variants of AoII, binary freshness is the most widely used metric when modeling  Markovian sources due to its simplicity and its direct relationship with the error probability of the estimates \cite{melih_BF_cache, melih_IF_CUS, melih_BF_Inf, melih_BF_gossip}. Even within the binary freshness metric, there have been a variety of versions of it, such as fresh when equal (FWE), fresh when close (FWC) and fresh when sampled (FWS), which incorporate various semantic aspects of the system as well \cite{nail_QS}. In this we work, we adopt the traditional FWE definition for binary freshness defined by,
\begin{align}
    \e[\Delta]=\limsup_{t_0\to\infty}\frac{1}{t_0}\e\left[\int_{0}^{t_0}\mathds{1}\{X(t)=\hat{X}(t)\}\,dt\right],
\end{align}
where $\mathds{1}\{\cdot\}$ is the indicator function, $\e[\Delta]$ is the average binary freshness, $X(t)$ is the state of the Markovian source and $\hat{X}(t)$ is its estimate at the remote monitor.

When analyzing the binary freshness metric, majority of prior works have utilized the martingale estimator due to its simplicity. However, this can be detrimental to the system performance from a freshness perspective, especially in pull-based update systems, where the remote estimator lack any knowledge related to the state of the Markovian source. For example, in the unfortunate event that we have sampled a less probable state, the martingale estimator is forced to retain this less probable state as its estimate until the next sampling instance. Replacing the martingale estimator with an MAP estimator is ideal, however can be analytically challenging in the continuous time domain making the design of sampling strategies an arduous task. Motivated by this fact, we introduce a new class of estimators called structured estimators which is an amalgamation between MAP and martingale estimators, utilizing the best from both worlds.

\begin{figure}
    \centering
    \includegraphics[width=\linewidth]{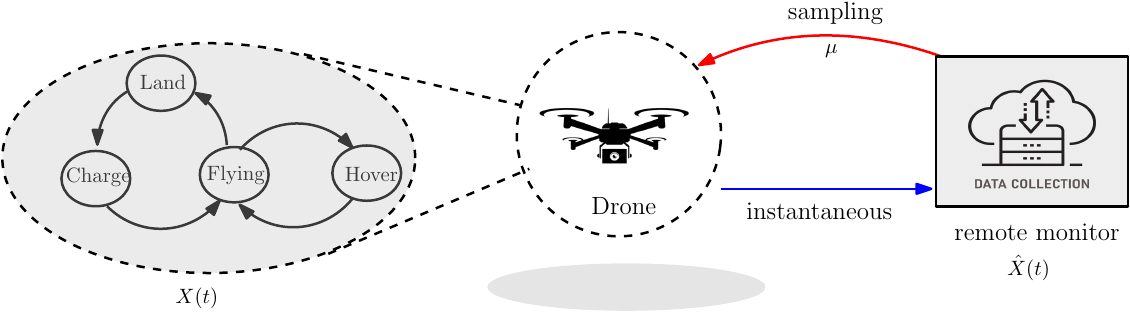}
    \caption{Query-based (pull-based) sampling of a Markovian source.}
    \label{fig:sys_model}
\end{figure}

In this work, we consider the problem of query-based sampling of a continuous time Markov chain (CTMC), where a remote monitor sends queries at exponential intervals to the Markov source, which instantaneously responds back with the system state to the remote monitor (see Fig.~\ref{fig:sys_model}). Our contributions can be summarized as follows: We introduce three new estimators, named exponential, Erlang and $\tau$-MAP estimators, and we provide closed-form analytical expressions for the binary freshness metric under each of them. Through theoretical and numerical results, we show that through a slight modification to the martingale estimators from which $\tau$-MAP estimators are derived, a significant gain in the binary freshness metric can be achieved. 

\section{Related Work}
The work in \cite{AoI_Markov} studies the problem of remote estimation of a discrete time Markovian source where they show that a sampling policy that minimizes the AoI does not necessarily minimize the error performance. This work was then extended to the AoII metric in \cite{AoII_Markov}, where the authors study the problem of minimizing AoII for a binary discrete time Markov source. There, the authors restrict their analysis to binary symmetric Markov chains with inter state transition probability $p<\frac{1}{2}$. The main reason for this restriction is that, when $p<\frac{1}{2}$, the martingale estimator coincides with the MAP estimator, making the analysis simpler. One of the few avenues in this domain, which considers a MAP estimator is the work done in \cite{ismail_map}. Here, the authors consider the problem of AoII minimization with the use of a MAP estimator for discrete time Markov sources. However, they resort to reinforcement learning techniques to circumnavigate some of the challenges posed by the MAP estimate. 

The work in \cite{Markov_machines} looks into the problem of query-based sampling of Markov machines where they look into the problem of maximizing binary fresh and other related metrics. However, they too resort to a martingale estimator which may lead to a crude decision process. The closest to our work is \cite{nail_QS}, where they look into the problem of rate allocation for query based sampling of multiple heterogeneous CTMCs with the use of a martingale estimator.  In this work, we improve upon work in \cite{nail_QS} with the use of the structured estimators, however, we restrict our analysis to monitoring a single CTMC and leave the sampling rate allocation problem for multiple CTMCs for future work.

\section{Preliminaries} \label{sec:prelim}
Let $X(t)$ be a finite state, irreducible CTMC with $S$ states where $X(t)\in \{1,2,\dots,S\}$. Let $Q=\{q_{ij}\}_{i,j\in S}$ be the generator matrix of this CTMC where $q_{ii}=-\sum_{j\neq i}q_{ij}=-q_i$. Therefore, $Q\mathbf{1}=0$, where $\mathbf{1}$ is a column vector of all ones.  Since $X(t)$ is irreducible and has a finite state space, its stationary distribution denoted by the column vector $\bm{\pi}=\{\pi_1,\pi_2,\dots,\pi_S\}$ exists and satisfies $\bm{\pi}^TQ=0$. Moreover, the transition matrix $P(t)=e^{Qt}$ of the CTMC will converge $\mathbf{1}\bm{\pi}^T$ as $t\to \infty$ (ergodic). Hence, we have the following intriguing property as a consequence of the ergodicity of the CTMC.

\begin{lemma}\label{lem:T}
    If the $i^*=\argmax_{i\in S}\pi_i$ is unique, then  $\exists~\tau^*<\infty$, such that the $\argmax_{i \in S} \bm{v}_j^TP(t)=i^*$ for all $t>\tau^*$ and $\forall j\in S$, where $\bm{v}_j$ is a vector of all zeros except for an one at the $j$th index.
\end{lemma}

The proof of Lemma \ref{lem:T} is given in the Appendix \ref{appen:lem_T}. In essence, Lemma \ref{lem:T} states that regardless of the starting state, the MAP estimate of $X(t)$ will be the same beyond a finite time $\tau^*$. We call these CTMCs to have a unique stationary maximum (i.e., $i^*=\argmax_{i\in S}\pi_i$ is unique).

\begin{definition}
    A CTMC is reversible with respect to measure $\bm{\pi}$ if it satisfies the detailed balance equations given by, $\pi_iq_{ij}=\pi_j q_{ji}$ $\forall i,j\in S$ and $i\neq j$.
\end{definition}

Throughout the paper, we simply say that a CTMC is time reversible, if the CTMC is reversible with respect to its stationary distribution $\bm{\pi}$. Let $\Pi$ be a diagonal matrix with $\Pi_{ii}=\pi_i$. If the CTMC is time reversible, $\Pi^{\frac{1}{2}}Q\Pi^{-\frac{1}{2}}$ is symmetric and hence can be diagonalized as $UDU^T$ where $D$ is a diagonal matrix whose diagonal elements are the eigenvalues of $Q$ given by $\{-d_1,-d_2,\dots,-d_S\}$. Since $Q\mathbf{1}=0$, there is $i_0\in S$ such that $d_{i_0}=0$. Moreover, we have that $\forall i\neq i_0, d_i>0$ \cite{gallager}. Thus, $cI-Q^T$, where $I$ is the identity matrix and $c\in\mathbb{R}$, is invertible $\forall c>0$.  Furthermore, we have that $\bm{u}_i^T\bm{u}_j=0, ~\forall i,j \in S$ and $i \neq j$, where $\bm{u}_i=\{u_{i1},u_{i2},\dots, u_{iS}\}$ is the $i$th column of the matrix $U$.

Let $R(t)$ denote an absorbing CTMC with states $\{1,2,\dots,\Gamma,\Gamma+1\}$ whose state transitions involve only transition from the $i$th state to the $(i+1)$th state with rate $\lambda$ where state $\Gamma+1$ is the absorbing state. Let $\tau^{(\lambda)}$ denote the sum of $\Gamma\in \mathbb{N}$, independent exponential random variables with rate $\lambda$. Then, $\tau^{(\lambda)}$ follows an Erlang distribution with shape parameter $\Gamma$ and rate $\lambda$. Moreover, $\tau^{(\lambda)}$ models the time to reach the absorbing state starting from state 1. Additionally, we have that $\e[\tau^{(\lambda)}]=\frac{\Gamma}{\lambda}$ and $\text{Var}(\tau^{(\lambda)})=\frac{\Gamma}{\lambda^2}$. Now consider, the random variable $\tau^{(\lambda \Gamma)}$. We have $\e[\tau^{(\lambda\Gamma)}]=\frac{1}{\lambda}$ and $\text{Var}[\tau^{(\lambda \Gamma)}]=\frac{1}{\lambda^2\Gamma}$. Therefore, $\tau^{(\lambda\Gamma)}$ converges to the deterministic value $\frac{1}{\lambda}$ in the mean square sense, and hence, in distribution as $\Gamma\to\infty$. 

\section{Structured Estimators}\label{sec:sys}
In this section, we present  the system model and various estimators considered in this work in addition to the commonly used martingale estimator in the literature. Let $X(t)$ be a finite, irreducible CTMC with a unique stationary maximum as described in Section \ref{sec:prelim}. This CTMC is sampled by a remote monitor at exponential random variable time intervals with a rate of $\mu$. We assume that once sampled, the sample is transmitted instantaneously to the remote monitor. Further, we assume that the initial state of our estimate at the receiver is randomly chosen according to the distribution $\bm{\pi}$. Next, we describe the estimators considered in this work.

\subsection{Martingale Estimator}
 Let $\hat{X}_M(t)$ denote the martingale estimator at the monitor. In here, $\hat{X}_M(t)$ will retain the previously received state as its estimate until updated by a new sample. Let $G(t)$ denote the last time the monitor received an update. Then, we define $\hat{X}_M(t)$ as follows,
\begin{align}
    \hat{X}_M(t)=X(G(t)).
\end{align}

\subsection{Exponential Estimator}\label{sec:exp_est}
Denote by $\hat{X}_{e,\lambda}(t)$ the exponential estimator. In here, once the monitor receives a new sample, it will start an exponential clock whose rate is $\lambda$. The monitor will retain the received sample as its estimate until a new sample is obtained or until the exponential timer runs out. If the exponential timer runs out, it will change its estimate to $i^*$,  which is the MAP estimate of $X(t)$ if we did not receive an update for a time period beyond $\tau^*$. It will then retain this $i^*$ state until a new sample is obtained. Each time a new sample is obtained, the exponential timer is reset. Let $S_i$ denote the $i$th sampling time. Since we sample at exponential intervals, we have $(S_{i+1}-S_i)\sim Exp(\mu)$. Let $\tilde{S}_i\sim Exp(\lambda)$ denote the realization of the exponential clock for the $i$th sample. Then, $\hat{X}_{e,\lambda}(t)$ can be defined as follows,
\begin{align}
    \hat{X}_{e,\lambda}(t)=\begin{cases}
        X(G(t)),& \text{if}~ t\in (S_i,\min\{S_i+\tilde{S}_i,S_{i+1}\}],\\
        i^*, & \text{if}~t\in (\min\{S_i+\tilde{S}_i,S_{i+1}\},S_{i+1}].
    \end{cases}
\end{align}

\subsection{Erlang Estimator}
\begin{figure}
    \centering
    \includegraphics[width=\linewidth]{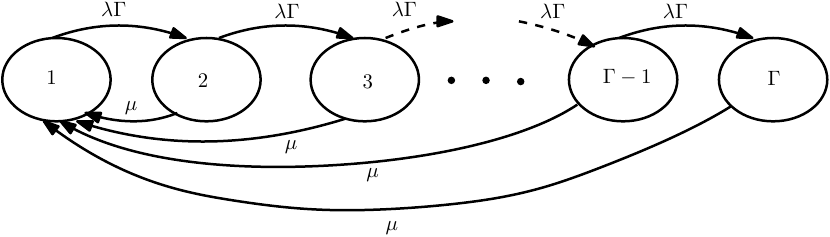}
    \caption{State transition diagram for $Y(t)$.}
    \label{fig:erlang_chain}
\end{figure}

Let $Y(t)$ be a CTMC independent of $X(t)$, with $\Gamma$ states where the state transitions occur only from the $k$th state to the $(k+1)$th state with rate $\lambda \Gamma$ and from the $k$th state to state $1$ with rate $\mu$. State $\Gamma$ can only transition to state $1$ with rate $\mu$; see Fig.~\ref{fig:erlang_chain}. Starting with $Y(0)=1$, the time to reach state $\Gamma$ when  $\mu=0$ is governed by an Erlang distribution whose rate and shape parameters are $\lambda \Gamma$ and $\Gamma-1$. Denote by $\hat{X}_{\Gamma,\lambda}(t)$, the Erlang estimator. Then, $\hat{X}_{\Gamma,\lambda}(t)$ is defined as follows,
\begin{align}
    \hat{X}_{\Gamma,\lambda}(t)=\begin{cases}
        X(G(t)),& \text{if}~Y(t)\neq \Gamma,\\
        i^*,&\text{if}~Y(t)=\Gamma.
    \end{cases}
\end{align}

\subsection{$\tau$-MAP Estimator}
Let $\hat{X}_\tau(t)$ denote the $\tau$-MAP estimator and $\delta(t)=t-G(t)$, the time elapsed since the last sample. In here, the monitor retains the previous sample until a new sample is obtained or until $\delta(t)>\tau$. If $\delta(t)>\tau$ then it will change its estimate to $i^*$. $\hat{X}_\tau(t)$ is defined as follows,
\begin{align}
    \hat{X}_\tau(t)=\begin{cases}
        X(G(t)),& \text{if}~ \delta(t)\leq \tau,\\
        i^*,& \text{if}~ \delta(t)> \tau.
    \end{cases}
\end{align}

\begin{remark}\label{rem:convergence}
    As $\Gamma \to \infty$, $\hat{X}_{\Gamma,\frac{1}{\tau}}(t)$ converges to  $\hat{X}_\tau(t)$ in distribution.
\end{remark}

\section{Main Results}
Now, we present our analytical results for the binary freshness metric under the estimators described in Section \ref{sec:sys}. Let $\text{tr}(\cdot)$ be the trace operator and denote by $\e[\Delta_{M}]$, $\e[\Delta_{e,\lambda}]$, $\e[\Delta_{\Gamma,\lambda}]$ and $\e[\Delta_{\tau}]$ the binary freshness obtained under the martingale, exponential, Erlang and $\tau$-MAP estimators, respectively. In here, we will only provide analytical results for the latter three estimators. The analysis for the martingale estimator can be found in \cite{nail_QS}. 

\begin{theorem}\label{thrm:exp_BF}
    Under an exponential estimator, $\e[\Delta_{e,\lambda}]$ is given by,
    \begin{align}
       \e[\Delta_{e,\lambda}]= \mu\text{tr}\left(\tilde{Q}_0\Pi\right)+\frac{\lambda}{\mu+\lambda}\pi_{i^*},\label{eqn:exp_BF}
    \end{align}
 where $\tilde{Q}_0=\left((\mu+\lambda)I-Q^T\right)^{-1}$.
\end{theorem}

The proof of Theorem \ref{thrm:exp_BF} follows similar arguments to that of the Erlang estimator, and hence its proof is omitted, and Erlang estimator proof is given in detail.

\begin{remark}
  $\e[\Delta_{e,0}]$ simplifies to the exact expression for the binary freshness metric under a martingale estimator obtained in \cite{nail_QS}. 
\end{remark}

\begin{theorem}\label{thrm:erlang_BF}
    Under an Erlang estimator, $\e[\Delta_{\Gamma,\lambda}]$ is given by,
    \begin{align}
       \e[\Delta_{\Gamma,\lambda}]= \mu\sum_{k=1}^{\Gamma-1}(\lambda \Gamma)^{k-1}\text{tr}\left(\tilde{Q}^{k}\Pi\right)+\frac{(\lambda \Gamma)^{\Gamma-1}}{(\mu+\lambda \Gamma)^{\Gamma-1}}\pi_{i^*},\label{eqn:erlang_BF}
    \end{align}
 where $\tilde{Q}=\left((\mu+\lambda \Gamma)I-Q^T\right)^{-1}$.
\end{theorem}

The proof of Theorem \ref{thrm:erlang_BF} can be found in the Appendix \ref{appen:erlang_BF}. Next, we analyze the expression in Theorem \ref{thrm:erlang_BF} explicitly for time reversible CTMCs.

\begin{corollary}\label{cor:erl_tr}
    If $X(t)$ is time reversible, then under an Erlang estimator, $\e[\Delta_{\Gamma,\lambda}]$ is given by,
    \begin{align}
        \e[\Delta_{\Gamma,\lambda}]=&\sum_{i=1}^S\frac{a_i\mu}{d_i+\mu}\left( 1-\left(\frac{\lambda\Gamma}{d_i+\mu+\lambda\Gamma}\right)^{\Gamma-1}\right)\nonumber\\
        &\qquad+\frac{(\lambda \Gamma)^{\Gamma-1}}{(\mu+\lambda \Gamma)^{\Gamma-1}}\pi_{i^*},\label{eqn:TR_BF_K}
    \end{align}
    where $-d_i$ are the eigenvalues of $Q$ and $a_i=\sum_{j=1}^S\pi_ju_{ij}^2$.
\end{corollary}

The proof of Corollary \ref{cor:erl_tr} is given in Appendix \ref{appen:cor_erl_tr}. Now, from \eqref{eqn:exp_BF} and \eqref{eqn:erlang_BF}, note that $\e[\Delta_{e,\lambda}]=\e[\Delta_{2,\frac{\lambda}{2}}]$. Therefore, by setting $\Gamma=2$ and $\lambda=\frac{\lambda}{2}$ in \eqref{eqn:TR_BF_K}, we can obtained the corresponding counterpart of Corollary \ref{cor:erl_tr} for exponential estimators.

\begin{corollary}
    If $X(t)$ is time reversible, then under an exponential estimator, $\e[\Delta_{e,\lambda}]$ is given by,
    \begin{align}
        \e[\Delta_{e,\lambda}]=\sum_{i=1}^S\frac{a_i\mu}{d_i+\mu+\lambda}+\frac{\lambda}{\mu+\lambda}\pi_{i^*}.
    \end{align}
\end{corollary}

By Remark \ref{rem:convergence} and dominated convergence \cite{koralov_sinai}, we have that $\e[\Delta_\tau]=\e[\lim_{\Gamma\to\infty}\Delta_{\Gamma,\frac{1}{\tau}}]=\lim_{\Gamma \to \infty}\e[\Delta_{\Gamma,\frac{1}{\tau}}]$. This is valid for any generic CTMC with a unique stationary maximum. 

Next, from Corollary \ref{cor:erl_tr} and using the fact that $\lim_{n\to\infty}\left(1+\frac{1}{n}\right)^n=e$, we obtain the expression for $\e[\Delta_\tau]$ for time reversible CTMCs.

\begin{corollary}
    If $X(t)$ is time reversible, then under an $\tau$-MAP, $\e[\Delta_{\tau}]$ is given by,
    \begin{align}
    \e[\Delta_{\tau}]=\sum_{i=1}^S\frac{a_i\mu}{d_i+\mu}(1-e^{-(d_i+\mu)\tau})+e^{-\mu \tau}\pi_{i^*}.
    \end{align}
\end{corollary}

Next, we present an important relationship between the martingale and $\tau$-MAP estimators. The proof of Theorem \ref{thrm:mart_v_map} can be found in Appendix \ref{appen:mart_v_map}.
\begin{theorem}\label{thrm:mart_v_map}
   Let $\tau^*$ be as defined in Lemma \ref{lem:T}. Then, the following relation holds: $\e[\Delta_M] \leq \e[\Delta_{\tau^*}]$.
\end{theorem}

\begin{figure}[!t]
    \centering
    \includegraphics[width=0.75\linewidth]{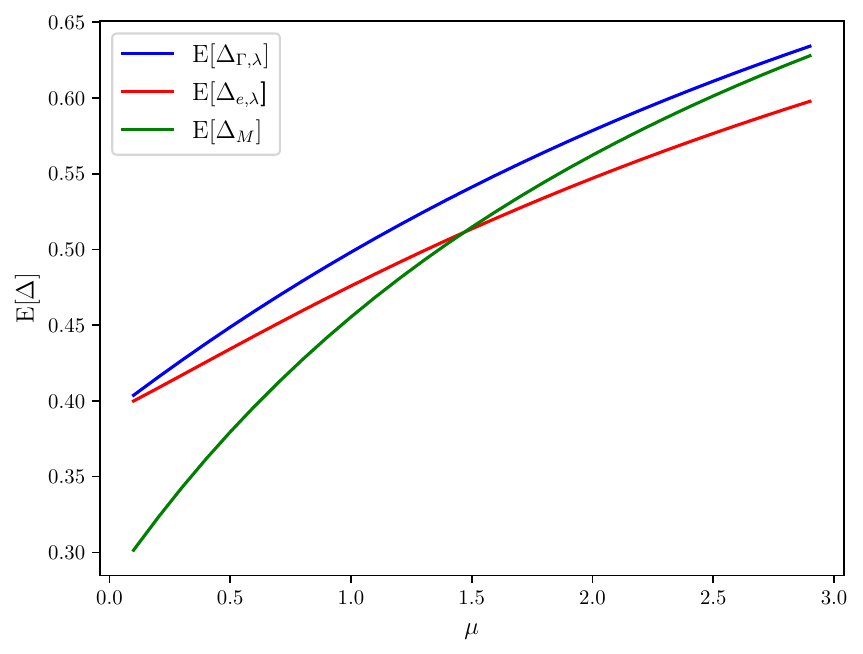}
    \caption{Variation of binary freshness with the sampling rate for $\Gamma=10$ and $\lambda=\frac{1}{\tau^*}$ for a generic 4-state CTMC.}
    \label{fig:var_mu}
\end{figure}

\section{Numerical Results}
In this section, we compare and evaluate the performance of the structured estimators as we change their respective parameters. First, we will compare how the binary freshness varies under the martingale, exponential and Erlang estimators, as we change the sampling rate for a generic CTMC of four states. In all numerical experiments we only consider CTMCs with a unique stationary maximum. As illustrated in Fig.~\ref{fig:var_mu}, the exponential estimator (red) outperforms the martingale estimator (green) at low sampling rates while the opposite behavior is observed at high sampling rates. This is due to the fact that, for high sampling rates, when the exponential timer runs out before we take the next sample, it would most probably run out at smaller time intervals (smaller than $\tau^*$). Thus, our estimator will shift to $i^*$ prematurely decreasing the binary freshness metric. At all sampling rates, we can see that the Erlang estimator outperforms the other two estimators. For higher sampling rates the performance of the martingale estimator approaches that of the Erlang estimator, while for lower sampling rates the exponential estimator exhibits comparable performance to the Erlang estimator.

\begin{figure}[t]
    \centering
    \includegraphics[width=0.75\linewidth]{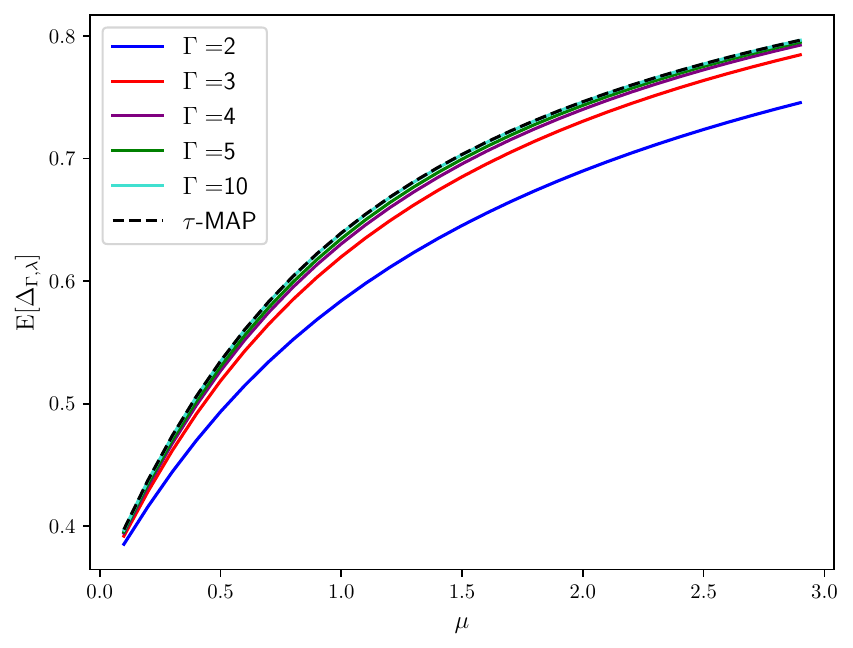}
    \caption{Variation of binary freshness with the sampling rate for different $\Gamma$ values with $\lambda=\frac{1}{\tau^*}$ and $\tau=\tau^*$ for a time reversible 5-state CTMC.}
    \label{fig:var_gamma_mu}
\end{figure}

\begin{figure}[t]
    \centering
    \includegraphics[width=0.75\linewidth]{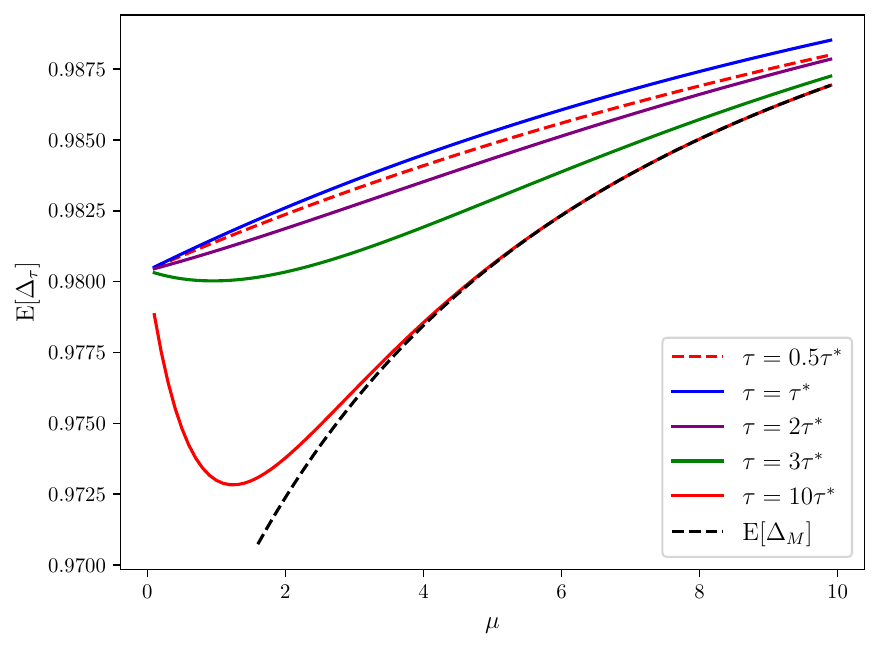}
    \caption{Variation of binary freshness with the sampling rate for different $\tau$ under $\tau$-MAP estimator for a 2-state CTMC.}
    \label{fig:var_mu_T}
\end{figure}

Next, we evaluate how the freshness varies as we increase $\Gamma$ of the Erlang estimator for a time reversible CTMC with $5$ states. As depicted in Fig.~\ref{fig:var_gamma_mu}, higher $\Gamma$ values lead to better binary freshness. Moreover, the curves converge and approache the $\tau$-MAP curve as $\Gamma$ increases. In the next experiment, we give an example which demonstrates a scenario where higher sampling can have an adverse effect on freshness if the correct $\tau$ value is not selected in the $\tau$-MAP estimator. For this particular example we consider a two state CTMC (time reversible) whose rate transitions are $q_{12}=5$ and $q_{21}=0.1$. Thus, we have $i^*=2$. As seen in Fig.~\ref{fig:var_mu_T}, the highest freshness is achieved when $\tau=\tau^*$ whereas deviating from $\tau^*$ seems to degrade the performance. It is worth noting that if $\tau$ is too high, then as we increase $\mu$, the freshness first decreases and then starts to increase. This is due to the fact, if $\tau$ is too large and $\mu$ is small, if we sample a less probable state, then it would take a long time for the estimator to either shift to the MAP state or for a new sample to be obtained, hence degrading the performance.

\section{Conclusion}
In this work, we introduced three new estimators termed as exponential, Erlang and $\tau$-MAP estimators, whose goal is to incorporate certain characteristics of the MAP estimate while still being analytically tractable. Through theoretical and numerical results, we show that $\tau$-MAP estimators can have a significant gain compared to martingale estimators in terms of binary freshness. The $\tau$-MAP estimator can be easily adapted to the rate allocation problem when monitoring several heterogeneous CTMCs and is an interesting extension of this work. Further, modifying the $\tau$-MAP estimator to have state-dependent thresholds for $\tau$ is another interesting line of research. Additionally, the inclusion of multiple intermediate states before transitioning to state $i^*$ will further bring the $\tau$-MAP estimator closer to the MAP estimator.

\appendices
\section{Proof of Lemma \ref{lem:T}}\label{appen:lem_T}
Let $\tilde{i}=\argmax_{i\in S,i\neq i^*}\pi_i$ and $\epsilon=\frac{\pi_{i^*}-\pi_{\tilde{i}}}{4}>0$. Since $X(t)$ is ergodic, we have that $\lim_{t\to\infty}P(t)=\mathbf{1}\bm{\pi}^T$ which is equivalent to $\lim_{t\to\infty}\|P(t)-\mathbf{1}\bm{\pi}^T\|_{F}=0$ where $\|\cdot\|_F$ is the Frobenius norm. Therefore, $\exists~\tau^*>0$ such that $\forall t>\tau^*$, $\|P(t)-\mathbf{1}\bm{\pi}^T\|_{F}<\sqrt{\epsilon}$. Now, for the rest of the proof, consider that $t>\tau^*$. This implies that $|P_{ij}(t)-\pi_j|<\epsilon$. Then, for $j\neq i^*$, the following relations hold,
\begin{align}
    P_{ii^*}-P_{ij}&>(\pi_{i^*}-\epsilon)-(\pi_{j}+\epsilon)\\
    &>\pi_{i^*}-\pi_{\tilde{i}}-2\epsilon=\frac{\pi_{i^*}-\pi_{\tilde{i}}}{2}>0.
\end{align}
Therefore, for $t>\tau^*$, we have that $\argmax_{i\in S}\bm{v}_j^TP(t)=i^*$.
\section{Proof of Theorem \ref{thrm:erlang_BF}}\label{appen:erlang_BF}
Let $Z(t)=(X(t),\hat{X}(t),Y(t))$. Then, $Z(t)$ is an irreducible finite state MC. Let $\bm{\Phi}=\{\phi_{i,j,k}\}_{i,j,k \in S}$ be the stationary distribution of $Z(t)$. Then, the global balanced equations yield the following,
\begin{itemize}
    \item $1<k<\Gamma$
    \begin{align}
    (q_i+\mu+\lambda \Gamma)\phi_{i,j,k}=\lambda \Gamma\phi_{i,j,k-1}+\sum_{l\neq i}\phi_{l,j,k}q_{li}.
    \end{align}
    \item $i \neq j$, $k=1$
    \begin{align}
      (q_i+\mu+\lambda \Gamma)\phi_{i,j,1}=\sum_{l\neq i}\phi_{l,j,1}q_{li}.
    \end{align}
    \item  $i=j$, $k=1$
    \begin{align}
      (q_i+\lambda \Gamma)\phi_{i,i,1}=&\sum_{l\neq i}\phi_{l,i,1}q_{li}+\mu\sum_{m>1}\sum_l\phi_{i,l,m}\nonumber\\
      &+\mu\sum_{l\neq i}\phi_{i,l,1}.
    \end{align}
    \item $k=\Gamma$
    \begin{align}
     (q_i+\mu)\phi_{i,i^*,\Gamma}=\lambda \Gamma \sum_{l}\phi_{i,l,\Gamma-1}+\sum_{l\neq i} \phi_{l,i^*,\Gamma}q_{li}.
    \end{align}
\end{itemize}
Simple rearrangement of the above equations along with the facts that $\sum_{l,m}\phi_{i,l,m}=\pi_i$ and $q_{ii}=-q_i$, gives THE following set of equations
\begin{itemize}
    \item $1<k<\Gamma$
    \begin{align}
    (\mu+\lambda \Gamma)\phi_{i,j,k}=\lambda \Gamma\phi_{i,j,k-1}+\sum_{l}\phi_{l,j,k}q_{li}.
    \end{align}
    \item $i \neq j$, $k=1$
    \begin{align}
      (\mu+\lambda \Gamma)\phi_{i,j,1}=\sum_{l}\phi_{l,j,1}q_{li}.
    \end{align}
    \item  $i=j$, $k=1$
    \begin{align}
      (\mu+\lambda \Gamma)\phi_{i,i,1}&=\sum_{l}\phi_{l,i,1}q_{li}+\mu \pi_i.
    \end{align}
    \item $k=\Gamma$
    \begin{align}
     \mu\phi_{i,i^*,\Gamma}=\lambda \Gamma \sum_{l}\phi_{i,l,\Gamma-1}+\sum_{l} \phi_{l,i^*,\Gamma}q_{li}.
    \end{align}
\end{itemize}
Let $\Phi_k=\{\phi_{i,j,k}\}_{i,j\in S}$ for $1\leq k<\Gamma$, be matrices of size $S \times S$ and let $\Phi_{\Gamma}=\{\phi_{i,i^*,\Gamma}\}_{i\in S}$ be a column vector. Then, the above equations can be denoted in matrix notation as follows:
\begin{itemize}
    \item $k=1$
    \begin{align}
        (\mu+\lambda \Gamma)\Phi_1=Q^T\Phi_1+\mu \Pi.\label{eqn:mat_1}
    \end{align}
    \item $1\leq k<\Gamma$
    \begin{align}
         (\mu+\lambda \Gamma)\Phi_k=Q^T\Phi_k+\lambda \Gamma \Phi_{k-1}.\label{eqn:mat_k}
    \end{align}
    \item $k=\Gamma$
    \begin{align}
        \mu \Phi_\Gamma=\lambda \Gamma \Phi_{\Gamma-1}\mathbf{1}+Q^T\Phi_\Gamma.\label{eqn:col_K}
    \end{align}
\end{itemize}
Let $\tilde{Q}=\left((\mu+\lambda \Gamma)I-Q^T\right)^{-1}$. From \eqref{eqn:mat_1}, we have,
\begin{align}
    \Phi_1=\mu\tilde{Q}\Pi. \label{eqn:Phi_1}
\end{align}
From \eqref{eqn:Phi_1} and \eqref{eqn:mat_k}, we have for $1<k<\Gamma$,
\begin{align}
    \Phi_k=\mu(\lambda \Gamma)^{k-1}\tilde{Q}^k\Pi. \label{eqn:Phi_k}
\end{align}
Finally, from \eqref{eqn:col_K} and \eqref{eqn:Phi_k} we get,
\begin{align}
    \Phi_\Gamma&= \mu(\lambda \Gamma)^{\Gamma-1}\hat{Q}\tilde{Q}^{\Gamma-1}\Pi\mathbf{1}
          &=\mu(\lambda \Gamma)^{\Gamma-1}\hat{Q}\tilde{Q}^{\Gamma-1}\bm{\pi}.\label{eqn:col_Phi_K}
\end{align}
where $\hat{Q}=\left(\mu I-Q^T\right)^{-1}$.
Note that since $\bm{\pi}^TQ=0$, we have that $\bm{\pi}$ is an eigenvector of $\tilde{Q}$ and $\hat{Q}$ with eigenvalues $\frac{1}{(\mu+\lambda \Gamma)}$ and $\frac{1}{\mu}$, respectively. Hence, \eqref{eqn:col_Phi_K} can be further simplified to the following,
\begin{align}
    \Phi_\Gamma= \frac{(\lambda \Gamma)^{\Gamma-1}}{(\mu+\lambda \Gamma)^{\Gamma-1}}\bm{\pi}.
\end{align}
Now, $\e[\Delta_{\Gamma,\lambda}]$ can be found as follows,
\begin{align}
    \e[\Delta_{\Gamma,\lambda}]&=\sum_{k=1}^{\Gamma-1}\sum_{i}\phi_{i,i,k}+\phi_{i^*,i^*,\Gamma}\\
    &=\sum_{k=1}^{\Gamma-1}\text{tr}\left(\Phi_k\right)+\bm{v}_{i^*}^T\Phi_\Gamma\\
    &=\sum_{k=1}^{\Gamma-1}\text{tr}\left(\Phi_k\right)+\bm{v}_{i^*}^T\frac{(\lambda \Gamma)^{\Gamma-1}}{(\mu+\lambda \Gamma)^{\Gamma-1}}\bm{\pi}\\
    &=\sum_{k=1}^{\Gamma-1}\text{tr}\left(\Phi_k\right)+\frac{(\lambda \Gamma)^{\Gamma-1}}{(\mu+\lambda \Gamma)^{\Gamma-1}}\pi_{i^*}.
\end{align}

\section{Proof of Corollary \ref{cor:erl_tr}}\label{appen:cor_erl_tr}
Since $Q$ is time reversible, we have that $Q=\Pi^{-\frac{1}{2}}UDU^T\Pi^{\frac{1}{2}}$. This gives us $\tilde{Q}=\Pi^{\frac{1}{2}}U\tilde{D}U^T\Pi^{-\frac{1}{2}}$ where $\tilde{D}$ is a diagonal matrix with $\tilde{D}_{ii}=\frac{1}{d_i+\mu+\lambda\Gamma}$. Therefore, we have $\tilde{Q}^k= \Pi^{\frac{1}{2}}U\tilde{D}^kU^T\Pi^{-\frac{1}{2}}$. Hence, we have,
\begin{align}
    \text{tr}\left(\tilde{Q}^k\Pi\right)&=\text{tr}\left(\Pi^{\frac{1}{2}}U\tilde{D}^kU^T\Pi^{-\frac{1}{2}}\Pi\right)\\
    &=\text{tr}\left(U\tilde{D}^kU^T\Pi\right)\\
    &=\text{tr}\left(\sum_{i=1}^S\frac{1}{(d_i+\mu+\lambda\Gamma)^k}\bm{u}_i\bm{u}_i^T\Pi\right)\\
    &=\sum_{i=1}^S\frac{1}{(d_i+\mu+\lambda\Gamma)^k}\text{tr}\left(\bm{u}_i\bm{u}_i^T\Pi\right)\\
    &=\sum_{i=1}^S\frac{1}{(d_i+\mu+\lambda\Gamma)^k}\sum_{j=1}^S\pi_ju_{i,j}^2\\
    &=\sum_{i=1}^S\frac{a_i}{(d_i+\mu+\lambda\Gamma)^k}.
\end{align}
Then, substituting for $\text{tr}\left(\tilde{Q}^k\Pi\right)$ in \eqref{eqn:erlang_BF} yields,
\begin{align}
    \e[\Delta_{\Gamma,\lambda}]=&\frac{\mu}{\lambda\Gamma}\sum_{k=1}^{\Gamma-1} \sum_{i=1}^Sa_i\left(\frac{\lambda\Gamma}{d_i+\mu+\lambda\Gamma}\right)^k+\frac{(\lambda \Gamma)^{\Gamma-1}}{(\mu+\lambda \Gamma)^{\Gamma-1}}\pi_{i^*}\nonumber\\
    =&\sum_{i=1}^Sa_i\frac{\mu}{\lambda\Gamma}\sum_{k=1}^{\Gamma-1} \left(\frac{\lambda\Gamma}{d_i+\mu+\lambda\Gamma}\right)^k+\frac{(\lambda \Gamma)^{\Gamma-1}}{(\mu+\lambda \Gamma)^{\Gamma-1}}\pi_{i^*}\nonumber\\
    =&\sum_{i=1}^Sa_i\frac{\mu}{d_i+\mu}\left( 1-\left(\frac{\lambda\Gamma}{d_i+\mu+\lambda\Gamma}\right)^{\Gamma-1}\right)\nonumber\\
    &+\frac{(\lambda \Gamma)^{\Gamma-1}}{(\mu+\lambda \Gamma)^{\Gamma-1}}\pi_{i^*}.
\end{align}
\section{Proof of Theorem \ref{thrm:mart_v_map}}\label{appen:mart_v_map}
Let $S_i$ be defined as in Section \ref{sec:exp_est}. Since the sampling process is regenerative, we have 
\begin{align}
    \e[\Delta_{\tau}]=\frac{\e[\int_0^{S_1}\mathds{1}\{X(t)=\hat{X}_{\tau}(t)\}\,dt]}{\e[S_1]},
\end{align}
where $X(0)=i$ with probability $\pi_i$. Now, note that for $t<\tau^*$, we have $\hat{X}_M(t)=\hat{X}_{\tau}(t)=X(0)$. Further, $\e[S_1]=\mu$. Therefore, we have
\begin{align}
   &\e\left[\int_0^{S_1}\mathds{1}\{X(t)=\hat{X}_{\tau}(t)\}\,dt\Bigg\lvert S_1,S_1>\tau^*,X(0)=i\right]\nonumber\\
   &=\int_0^{S_1}\e[\mathds{1}\{X(t)=\hat{X}_{\tau}(t)|X(0)=i\}]\,dt\\
   &=\int_0^{\tau^*}\e[\mathds{1}\{X(t)=i\}]\,dt+\int_{\tau^*}^{S_1}\e[\mathds{1}\{X(t)=i^*\}]\,dt\\
   &=\int_0^{\tau^*}\e[\mathds{1}\{X(t)=i\}]\,dt+\int_{\tau^*}^{S_1}P_{ii^*}(t)\,dt\\
   &\geq\int_0^{\tau^*}\e[\mathds{1}\{X(t)=i\}]\,dt+\int_{\tau^*}^{S_1}P_{ii}(t)\,dt\label{eqn:map_ineq}\\
   &=\int_0^{\tau^*}\e[\mathds{1}\{X(t)=i\}]\,dt+\int_{\tau^*}^{S_1}\e[\mathds{1}\{X(t)=i\}]\,dt\\
   &=\int_0^{S_1}\e[\mathds{1}\{X(t)=\hat{X}_{M}(t)|X(0)=i\}]\,dt\\
   &=\e\left[\int_0^{S_1}\mathds{1}\{X(t)=\hat{X}_{M}(t)\}\,dt\Bigg\lvert S_1,S_1>\tau^*,X(0)=i\right].
\end{align}
The inequality \eqref{eqn:map_ineq} follows from the fact that $i^*$ is the MAP estimator for $X(t)$ when $t>\tau^*$ as a consequence of Lemma \ref{lem:T}. Further, the interchange of the integral and the expectation is justified by Tonelli's theorem. From the above relation, we have that,
\begin{align}
    \e&\left[\int_0^{S_1}\mathds{1}\{X(t)=\hat{X}_{\tau}(t)\}\,dt\Bigg\lvert S_1>\tau^*\right]\nonumber\\
    &\geq \e\left[\int_0^{S_1}\mathds{1}\{X(t)=\hat{X}_{M}(t)\}\,dt\Bigg\lvert S_1>\tau^*\right].
\end{align}
Also, since $\hat{X}_M(t)=\hat{X}_{\tau}(t)$ for $t<\tau^*$, we further have that,
\begin{align}
     \e&\left[\int_0^{S_1}\mathds{1}\{X(t)=\hat{X}_{\tau}(t)\}\,dt\Bigg\lvert S_1\leq\tau^*\right]\nonumber\\
    &= \e\left[\int_0^{S_1}\mathds{1}\{X(t)=\hat{X}_{M}(t)\}\,dt\Bigg\lvert S_1\leq\tau^*\right].
\end{align}
This proves the desired result.
\bibliographystyle{unsrt}
\bibliography{refs}

\end{document}